\documentclass[a4paper,12pt]{article}
\usepackage{graphicx}
\usepackage{amsmath}
\usepackage[german,english]{babel}
\begin{document}
\sloppy
\frenchspacing
\title{\bf Modeling Supply Networks and Business Cycles as Unstable Transport Phenomena}
\author{Dirk Helbing\\[3mm] 
Institute for Economics and Traffic,\\ 
Dresden University of Technology,\\ Andreas-Schubert Str. 23, 01062 Dresden, Germany}
\maketitle
\begin{abstract}
Physical concepts developed to describe instabilities in traffic flows
can be generalized in a way that allows one to understand the well-known
instability of supply chains (the so-called ``bullwhip effect'').
That is, small variations in the consumption rate can cause
large variations in the production rate of companies generating
the requested product. Interestingly, the
resulting oscillations have characteristic frequencies
which are considerably lower than the variations in the
consumption rate. This suggests that instabilities of
supply chains may be the reason for the existence of business
cycles. At the same time, we establish some link to queuing theory and
between micro- and macroeconomics.
\end{abstract}


\section{Introduction}

Concepts from statistical physics and non-linear dynamics have been very successful in
discovering and explaining dynamical phenomena in traffic flows \cite{reviews}. Many of these phenomena
are based on mechanisms such as delayed adaptation to changing conditions and 
the competition for limited resources, which are relevant for other systems as well. This includes
pattern formation such as segregation in driven granular media \cite{segregation} and
lane formation in colloid physics \cite{colloid} or biological physics (pedestrians, ants) \cite{pedants}.
Other examples are clogging phenomena at bottlenecks in freeway traffic \cite{phase}, panicking pedestrian crowds
\cite{panic}, or granular media \cite{clogging}. In the following study, we will focus on the phenomenon of
stop-and-go traffic \cite{stopgo} and its analogies. 
\par
Recently, economists and traffic scientists have wondered, whether 
traffic dynamics has also implications for the stability and management of supply chains 
\cite{Daganzo,Armbruster} or for the dynamics of business cycles \cite{Witt}. To explain business cycles, many
theoretical concepts have been suggested over the decades, 
such as the Schumpeter clock  \cite{business}. These are usually based on
macroeconomic variables such as investment, income, consumption, public expenditure, or the
employment rate, and their interactions.  In contrast, Witt {\em et al.} \cite{Witt} have recently 
suggested to interpret business cycles as self-organization phenomenon due to a linear instability 
of production dynamics related to stop-and-go waves in traffic or driven many-particle systems. 
In order to illustrate their idea, they have transferred a continuous macroscopic
traffic model and re-interpreted the single terms and variables. 
\par
The author believes that this is a very promising approach to understand business cycles, but
instead of simply transferring macroscopic traffic models, suggests 
to derive equations for business cycles from first principles, which means to derive the dynamics on
the macroscopic level from microscopic interactions in production systems. This would also make some contribution 
to the goal of understanding macroeconomics based on micro\-economics (or, in a wider sense,
based on the ``elementary interactions'' of individuals, here: production managers). 
\par
In order to make some progress in this direction, we will generalize \cite{preprint} an idea suggested by
Daganzo to describe the dynamics of supply chains \cite{Daganzo}.
Like the work by Armbruster {\em et al.} \cite{Armbruster}, 
his approach is related to traffic models as well, but he focusses on models in discrete space in order 
to reflect the discreteness of successive production steps. In order to 
describe the non-linear dynamics of production processes or interrelated
economic sectors, we will have to generalize these ideas to complex supply networks. 
In Sec.~\ref{Sec1}, we will first discuss ``macroscopic''
business cycles in a sectorally structured economy and compare 
them with stop-and-go traffic. Afterwards, in Sec.~\ref{SEC2}, we will develop a more fine-grained,
``microscopic'' description of the management of dynamically interacting production units
and relate it to classical queueing theory (which mainly focusses on stochastic fluctuations 
of production processes in a stationary state). Later on, in Sec.~\ref{SEC3} we will construct a mathematical relation
between the microscopic and the macroscopic level of description, while some
further research directions and other potential applications are indicated in the outlook of Sec. \ref{Sec4}. 

\section{Modelling ``Macroscopic'' Supply Networks} \label{Sec1}

\subsection{Economic Production Sectors}

\begin{figure}[h]
\begin{center}
\includegraphics[width=10cm]{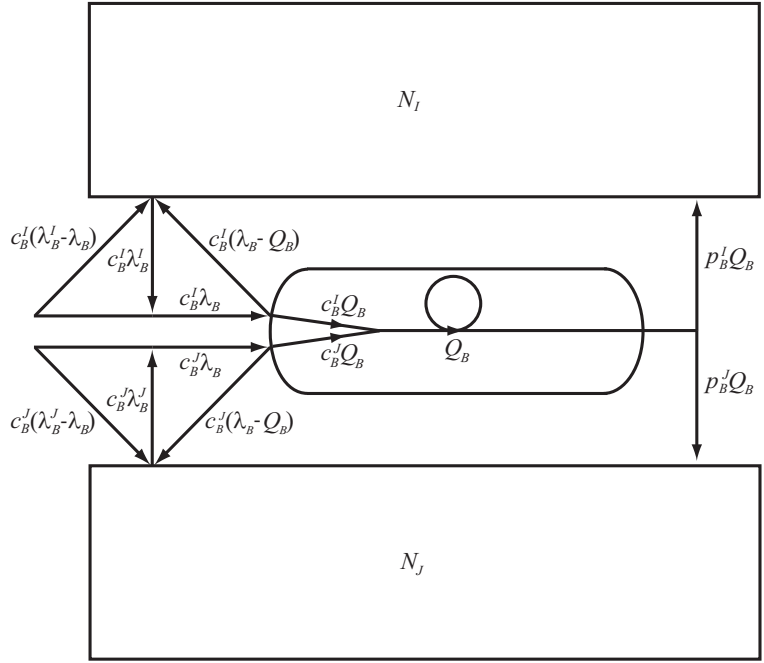}
\end{center}
\caption{Schematic illustration of the virtual and actual product flows for the case of two kinds of
products, $I$ and $J$. The picture displays
the various model variables related to production sector $B$, which is represented by the
oval structure in the center. Rectangles correspond to markets, arrows to
product flows.}
\label{figure1a}
\end{figure}

We will first investigate a simplified economic system with $U$ production sectors $B$ generating
certain kinds of products $I\in \{1,\dots,P\}$. The market for products $I$ as a function of time $t$ shall be
represented by the stock level (``inventory'') $N_I(t)$.
Let us assume that, in each production cycle, the production sector $B$ generates
$p_B^I$ products of kind $I$ and requires $c_B^J$ products of kind $J$ (``educts'') for production. If
$Q_B(t)$ is the number of production cycles per unit time, i.e.
a measure of the production rate, productivity or  ``throughput'' of sector $B$, the quantity of
products generated per unit time is $p_B^I Q_B(t)$, while the quantity of educts consumed per unit time
is $c_B^J Q_B(t)$ (see Fig.~\ref{figure1a}). 
The temporal change of the quantity of products $I$ in the market is, therefore, given by
the conservation equation
\begin{equation}
 \frac{dN_I}{dt} = \sum_B ( p_B^I - c_B^I) Q_B(t) \, .
\end{equation}
The production rate $Q_B(t)$ will be specified later on in Sec.~\ref{SEC2}. For the time
being, we will assume $Q_B(t)$ to agree with the actual feeding (``arrival'') rate $\lambda_B$
discussed in the next paragraph:
\begin{equation}
 \lambda_B(t) = Q_B(t) \, .
\end{equation}

\subsection{The Feeding Rates}

In the absence of capacity constraints, we would have the relation $\lambda_B = \rho_B^0 \mu_B$ for the 
time-dependent feeding
rate $\lambda_B$, where $\mu_B$ is the processing rate of sector $B$, i.e. the potential production rate
in the case of no inefficiencies. $\rho_B^0$ reflects the desired utilization. A value of $\rho_B^0 = 0.7$
is reasonable, as it guarantees a relatively high production rate at moderate and reliable
waiting times for finishing the products (see Sec.~\ref{SEC2a}). However, 
the actual utilization $\rho_B = \lambda_B / \mu_B$ does
not necessarily agree with the desired one, $\rho_B^0$. In order to illustrate this,
let us assume that $T_B^J$ is the average time needed to deliver to production sector~$B$
products of kind $J$, which are required for production (which will be called ``educts'' in the following).  
If $N_J$ is the quantity of actually
available products of kind $J$, the resulting maximum quantity of products that can be delivered per unit time is
$N_J /T_B^J$. As $c_B^J$ educts are required per production cycle, one cannot have a higher production rate 
than $N_J/ (T_B^J c_B^J)$ per unit time, which defines the maximum feeding rate
$\lambda_B^J$ of educt $J$ into production unit $B$:
\begin{equation}
 \lambda_B^J = N_J/ (T_B^J c_B^J) \, .
\end{equation}
Moreover, as the products can only be finished, if all required
educts are available in a production cycle, the actual feeding rate $\lambda_b$ is given by the
minimum of these values and of the desired feeding rate $\rho_B^0 \mu_B$ (see Fig.~\ref{figure1a}):
\begin{equation}
 \lambda_B(t) = 
   \min ( \rho_B^0 \mu_B  , \{ \lambda_B^J \} ) \, .
\label{minim}
\end{equation}
If $c_B^J = 0$, we set $\lambda_B^J \rightarrow \infty$, so that the corresponding term does not
have any impact on the value of the minimum function (\ref{minim}).
\par
Note that production has some analogy with chemical reactions, where one also requires a certain quantity
of educts to produce other (chemical) products. For chemical reactions in threedimensional space, however,
the reaction rate is given by multiplication of (a power of) the chemical concentrations of the educts,
i.e. the above minimum function is replaced by an algebraic product.
This difference is comparable to the Probabilistic AND and the Fuzzy AND in Fuzzy Logic,
which are applied when several conditions are to be met at the same time. The Probabilistic
AND is based on a multiplication of the logical values, while the Fuzzy AND corresponds
to their minimum.

\subsection{Adaptation of Production Speeds and Transport Capacities}

\begin{figure}[h]
\begin{center}
\includegraphics[height=10cm,angle=-90]{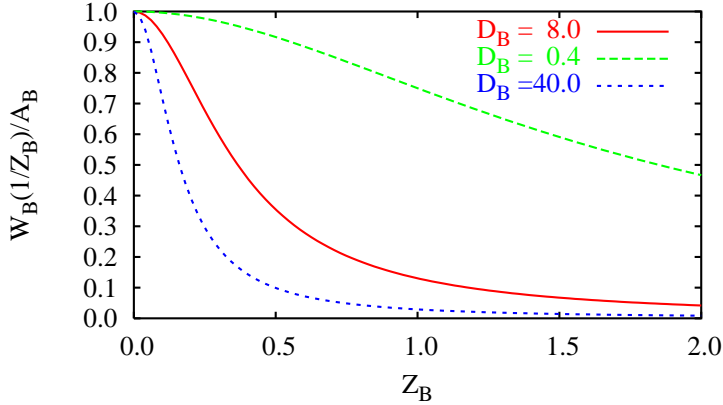} 
\end{center}
\caption{Steady-state production speed $W_B$ as a function of the scaled inventories $Z_B$ for
the parameter $B_B = 0.2$, arbitrary $A_B > 0$,  and various values of $D_B$. The production speed
decreases with increasing inventories. Throughout this paper we use the parameter values $B_B = 0.2$ and
$D_B = 8$.}
\label{FIG2}
\end{figure}

One important aspect is the adaptation of production to the time-dependent inventories $N_J(t)$. 
The adaptation of the desired utilization $\rho_B^0$ is delayed (see Sec.~\ref{SEC2}), 
and a change of the processing rate $\mu_B$, which requires an adaptation of production sectors 
or capacities, takes a considerable time as well. We will assume 
\begin{equation}
 \frac{d(\rho_B^0\mu_B)}{dt} = \frac{1}{\tau_B} \left[ 
 W_B\!\left( 1/Z_B, \dots \right) - \rho_B^0 \mu_B\right] 
\end{equation}
with a typical adaptation time $\tau_B$.
$W_B(1/Z_B,\dots)$ reflects the production rate of sector $B$ in steady state
as a function of the inventories $N_J$. For the time being, we will assume 
\begin{equation}
   \frac{1}{Z_B(t)} = \sum_J \frac{p_B^J X_B^J}{N_J(t)} \, ,
\end{equation}
i.e.~the production speed is adapted only to the inventories $N_J$ of the products $J$
generated by sector $B$, for which $p_B^J > 0$.
$X_B^J$ is an additional and constant prefactor.
This formula is defined in a way which gives always positive values
$Z_B(t)$, if not all coefficients $p_B^J$ are zero and the $N_J(t)$ stay positive (see below).
Normally, the production speed $W_B(1/Z_B,\dots)$ will increase with decreasing inventories
$N_J$, but it saturates
due to financial or technological limitations and inefficiencies. 
In this study, we will assume 
\begin{equation}
W_B(1/Z_B,\dots) = \max\left(A_B \frac{1+B_B Z_B}{1+B_B Z_B+D_B (Z_B)^2}, 0\right) \, ,
\end{equation}
where $Z_B$ could be called ``scaled inventories''
and $A_B$, $B_B$, and $D_B$ are suitable parameters (see Fig.~\ref{FIG2}).
\par
Furthermore, we will assume here that the transport capacities are adapted in parallel with the production speed
and represent the proportionality constants by $V_B^J$. This implies 
\begin{equation}
 \lambda_B^J = \rho_B^0 \mu_B V_B^J N_J/ c_B^J \, ,
\end{equation}
corresponding to the specification $1/T_B^J = \rho_B^0 \mu_B V_B^J$, which results in the
following formula for the feeding rates:
\begin{equation}
 \lambda_B(t) = \rho_B^0 \mu_B 
   \min \left(  1 , \left\{ \frac{V_B^J N_J}{c_B^J} \right\} \right) \, .
\end{equation}
Herein, the minimum extends over all indices $J$.

\subsection{Bull-Whip Effect and Stop-and-Go Traffic} \label{bull}

\begin{figure}[h]
\begin{center}
\includegraphics[height=10cm,angle=-90]{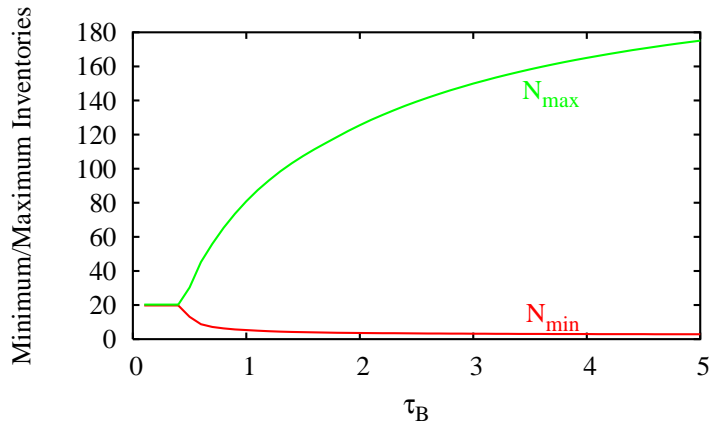} 
\end{center}
\caption{Minimum and maximum inventories as a function of the adaptation time $\tau_B$. The other model
parameters are specified as in Fig.~\ref{FIG4}, but the perturbation amplitude has been chosen
five times smaller. The difference between both curves is the amplitude of
the bullwhip effect, i.e. the time-dependent variation in the inventories. Note that there is a critical
adaptation time, below which perturbations are not amplified. In this case, the investigated
linear supply chain behaves stable.}
\label{FIG3}
\end{figure}

For a linear supply chain with $c_B^I = \delta_{B,I+1}$ and $p_B^I = \delta_{B,I}$
(where $\delta_{K,L} = 1$, if $K=L$, and $\delta_{K,L} = 0$ otherwise)
we obtain the particular set of equations 
\begin{equation}
\frac{dN_I}{dt} = Q_I(t) - Q_{I+1}(t)= \lambda_I(t) - \lambda_{I+1}(t)\, . 
\end{equation}
In the capacity-constrained case (characterized by small values of $V_B^J$), this leads to 
\begin{equation}
  \frac{dN_I}{dt} = V_I(t) N_{I-1}(t) - V_{I+1}(t) N_I(t)  
\label{eq2}
\end{equation}
with
\begin{equation}
  V_I(t) = \rho_I^0(t) \mu_I V_{I}^{I-1} 
\end{equation}
and
\begin{equation}
 \frac{d(\rho_I^0\mu_I)}{dt} = \frac{1}{\tau_I} \left[ 
 W_I\!\left( 1/Z_I\right) - \rho_I^0\mu_I \right] 
\label{eq0}
\end{equation}
or
\begin{equation}
 \frac{dV_I}{dt} = \frac{1}{\tau_I} \left[ V_{I}^{I-1} 
 W_I\!\left( 1/Z_I \right) - V_I(t) \right] \, .
\label{EQ0}
\end{equation}
Interestingly, Eqs. (\ref{eq2}) and (\ref{EQ0}) basically
agree with the macroscopic traffic flow model by Hilliges and Weidlich \cite{Hilliges},
where (\ref{eq2}) is analogous to the equation for the vehicle density and (\ref{EQ0}) corresponds to the
equation for the average vehicle speed $V_I$ in street segment $I$.
These equations behave linearly unstable with respect to perturbations of
$N_I(t)$, if $\tau_I$ exceeds a certain threshold (see Fig.~\ref{FIG3}), which depends on the maximum slope 
$W'(1/Z_I)$ of $W(1/Z_I)$ \cite{Hilliges}. As drivers (over-)react  with a time delay to a changing traffic situation in front,
stop-and-go traffic can emerge. The frequently observed
instability of supply chains, called the ``bull-whip effect'', occurs for similar reasons
(e.g. in the ``beer distribution game'' \cite{beer1,beer2}), see 
Fig.~\ref{FIG4}. 
\par
\begin{figure}[h]
\begin{center}
\includegraphics[height=10cm,angle=-90]{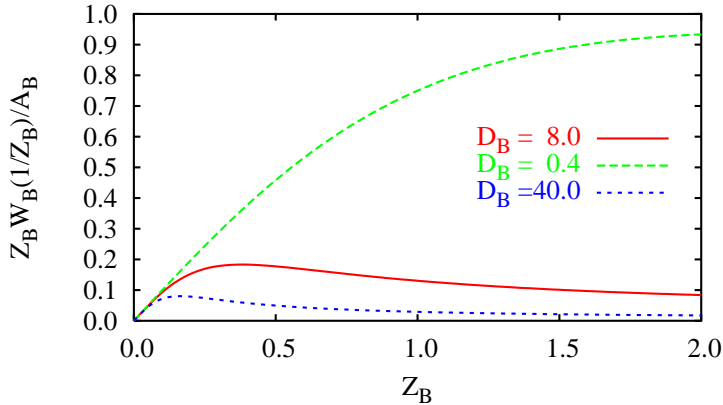} 
\end{center}
\caption{Steady-state throughput $Q_B$ as a function of the scaled inventories $Z_B$ for
the parameter $B_B = 0.2$, arbitrary $A_B$,  and various values of $D_B$ in the capacity-restricted
case of low transport rates $V_B^I$. For large enough values of $D_B$, the curve has a maximum 
at finite inventories. In this case, high steady-state flows $Q_B$
appear twice: for low (and linearly stable) inventories and higher (but potentially unstable) inventories
(see also Fig.~\ref{FIG8}).}
\label{FIG5}
\end{figure}

The mechanism behind this instability is the delay $\tau_I$ in the adaptation of the
production speeds and transport capacities, which implies an over- or under-production. The 
repeatedly or periodically resulting high inventories are 
due to temporary bottlenecks in the supply chain 
and could be avoided by appropriate control functions $W_B=W_B(\{N_J(t)\},\{dN_J/dt(t)\},\dots)$ \cite{Nagatani,Seba}.
Note that, instead of unstable production with high inventories and low production speeds,
one may reach the same average, but stable throughput at low inventories and high production speeds
(see Fig.~\ref{FIG5}). However, this is normally related with higher energy and maintainance costs, so that
production tends to operate in the linearly unstable regime.
\par
If the transport capacities are not a limiting factor (i.e. the parameters $V_B^J$ are large), 
instead of (\ref{eq2}) we have the set of equations
\begin{equation}
  \frac{dN_I}{dt} = \rho_I^0(t) \mu_I - \rho_{I+1}^0(t) \mu_{I+1} \, .
\end{equation}
Together with (\ref{eq0}), this basically corresponds to a particular microscopic traffic model, the
so-called optimal velocity model \cite{hyst}, if we restrict our comparison to the linear regime around the
steady state and identify $\rho_I^0 \mu_I$ with the actual velocity of vehicle $I$, but $N_I$ with the
inverse distance to the next car ahead (apart from proportionality constants). 
It is known \cite{Nagatani,hyst} that this model is linearly unstable, if
\begin{equation}
 V_{I}^{I-1} W'_I(1/Z_I) > \frac{1}{2\tau_I} \, , 
\end{equation}
i.e. if the adaptation time $\tau_I$ exceeds a certain threshold which 
depends on the slope $W'_I(1/Z_I)$ of $W_I(1/Z_I)$. 

\subsection{Business Cycles}

\begin{figure}[tbph]
\begin{center}
\includegraphics[height=10cm, angle=-90]{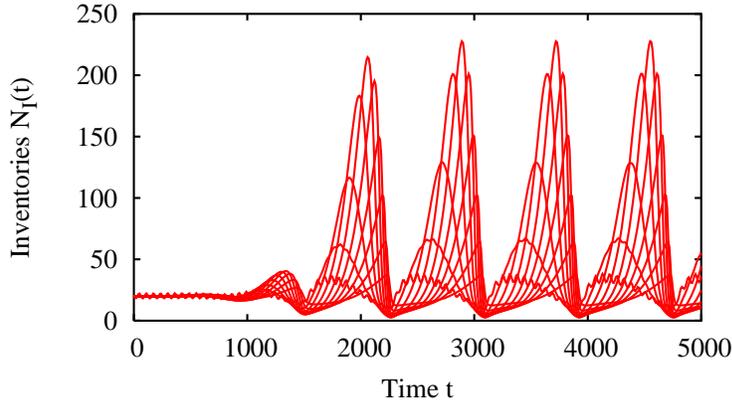} 
\end{center}
\caption{Variations of the inventories $N_I(t)$ of $U=P=10$ economic sectors $I$, triggered by a constant
consumption rate with a slight periodic perturbation, namely $V_{11}(t) = V_{11}^{10} W_I(1)
[1+ 0.1 \sin(0.1 t)]$.
One can clearly see the emergence of slow and synchronized oscillations of the inventories
$N_I(t)$, which are triggered
by small and fast perturbations in the consumption rate $V_{11}(t)$.
The unit time is one day and the model parameters are $A_B =A_I =100/V$, 
${V}_{I}^{I-1}= V =10^{-4}$, and $\tau_B = \tau_I = 90$. The initial and boundary conditions are
$N_I(0) = N_0(t) = 20 = X_I^I$, $V_I(0) = V W_I(X_I^I/N_I(0)) = V W_I(1)$, and $V_{11}(t)$ (see above).}
\label{FIG4}
\end{figure}

The most interesting point is the reaction of the system to a perturbation in the
throughputs $Q_B(t)$, for example a periodic perturbation of the so-called
consumption rate $Q_{U+1}(t)$ (see Secs.~\ref{Sec3} and \ref{SEC3d}). 
The resulting oscillations in the inventories can be 
much slower and are usually synchronized among different production sectors (see Figs.~\ref{FIG4}, 
\ref{Fig9}, and \ref{Fig10}). 
Therefore, they may explain business cycles as a self-organized phenomenon 
with slow dynamics on the time scale of several years. 
\par
\begin{figure}[h]
\begin{center}
\includegraphics[height=9.5cm,angle=-90]{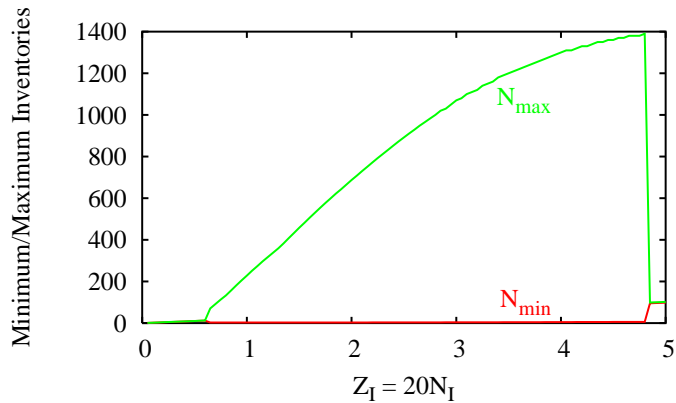} 
\end{center}
\caption{Minimum and maximum inventories as a function of the scaled steady-state inventory 
$Z_I(0)$. The other model
parameters are specified as in Fig.~\ref{FIG4}, but the perturbation amplitude has been chosen
five times smaller (of the order of 2\%). The difference between both curves reflects the amplitude of
the time-dependent variation in the inventories. Note that, in the capacity-constrained Hilliges-Weidlich case
investigated here, the change of the amplitude (and the associated kind of non-equilibrium
phase transition) is continuous. The supply chain is stable with respect to perturbations,
where both curves agree. This is the case at small and large steady-state inventories.}
\label{FIG8}
\end{figure}

The reduction in the resulting oscillation frequency compared with the perturbation frequency is
also known from stop-and-go traffic. It has been explained by the non-linearity in the
model equations. However, there is a significant difference between the dynamics of traffic flows and supply
chains: While stop-and-go waves have a characteristic amplitude independently of the average vehicle
density, the amplitude of oscillations in the inventories change continuously in
the capacity-constrained case (see Fig.~\ref{FIG8}). In other words, the phase transition
from stable to unstable traffic flow is hysteretic (i.e., of first order) \cite{Kern,Opus}, while the phase transition
from stable to unstable supply chains in the capacity-constrained case appears to be continuous (i.e., of second order).
This seems to be a particular property of the Hilliges-Weidlich model, while the optimal-velocity model
mentioned in Sec.~\ref{bull} is known to display a hysteretic transition \cite{hyst}. Therefore, a hysteretic transition
is found for supply chains which are not contrained by their transport capacities. The transition between these
two different regimes would be interesting to investigate.
\par
Note that, in order to find the emergence of slow oscillations, i.e. of business cycles,
we do not need to have a linear supply chain. Supply networks can display similar features (see Sec.~\ref{Topo}).
The  only requirement is that no stationary state exists or the stationary state is linearly unstable with respect
to perturbations. This is intensively studied for particular network types 
in a recent scientific collaboration \cite{Seba}.
\par
Summarizing our present insights, 
preconditions for the emergence of business cycles are
\begin{itemize}
\item large values of the adaptation times $\tau_B$,
\item significant changes of the production speeds $W_B$ with changing inventories,
\item the presence of perturbations, if a steady system state exists,
\item no consideration of forecasts or poor forecasts of the future time development of the 
inventories $N_I(t)$,
\item a highly non-linear interaction among the production sectors.
\end{itemize}
We should finally underline that the above considerations can be applied to economic systems 
with any number of sectors and any network structure. We do not even need to assume a sectorally
structured economy, as the same kinds of equations apply on the ``microscopic'' level of production networks,
only with a significantly higher number of equations (see Sec.~\ref{Sec2}). The special cases discussed above
have been chosen only for illustrative reasons.

\section{``Microscopic'' Model of Production Processes} \label{SEC2}

In this section, we will formulate a generalized model for the dynamical interaction of
production processes. Compared to the description of interacting economic sectors in the
previous section, this approach may be called microscopic. Note that the concept developed in
this section is required for two reasons: First in order to be able to describe real production processes, 
which involves buffers and other variables and calls for a more complex model.
Second in order to allow the derivation of the macroeconomic dynamics from 
microeconomic assumptions regarding the production management of single 
companies (see Sec.~\ref{SEC3}).
Readers not interested in these aspects may skip this section and continue 
with Sec.~\ref{Topo}.

\subsection{The Production Units in Terms of Queueing Theoretical Quantities}\label{SEC2a}

\begin{figure}[h]
\begin{center}
\includegraphics[width=10cm]{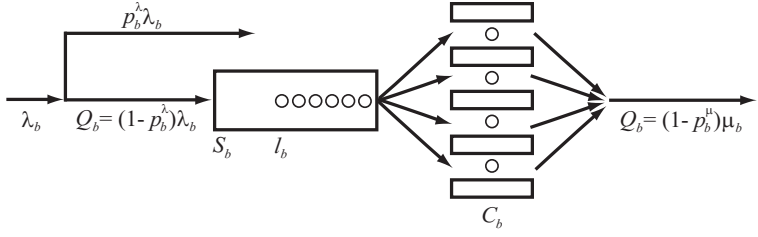}
\end{center}
\caption{Schematic 
illustration of a production unit $b$ as a queueing system with
a limited storage capacity $S_b$ and $C_b$ parallel production channels. The arrival rate
$\lambda_b$, the departure rate $\mu_b$, and effects of inefficiencies are indicated.}
\label{ADD}
\end{figure}

We will now investigate a system with $u$ production units (machines or factories) $b \in$ $\{1,2,\dots,u\}$ 
producing $p$ products $i, j\in \{1,2,\dots,p\}$. The respective production process is characterized by 
parameters $c_b^j$ and $p_b^i$: In each production cycle, production unit $b$ requires
$c_b^j$ products (``educts'') $j\in \{1,\dots,p\}$ and produces $p_b^i$ products $i \in \{1,\dots,p\}$.
The number of production cycles of production unit $b$ per unit time is a measure of
the throughput and shall be represented by $Q_b(t)$. For production in the steady state, we can
often approximate this quantity in terms of variables from queueing theory \cite{queuing}. For this, let
$\lambda_b$ be the feeding (``arrival'') rate, $C_b$ the number of parallel production channels, $\mu_b$
the overall processing (``departure'') rate (i.e. $C_b$ times the processing rate of a {\em single} 
production channel), 
\begin{equation}
 \rho_b = \lambda_b / \mu_b 
\label{util}
\end{equation}
the utilization, and
$S_b$ the storage capacity of production unit $b$. Then, the following relationships apply
(see Fig.~\ref{ADD}):
\begin{equation}
 Q_b(t) = (1-p_b^\lambda) \lambda_b(t) = (1-p_b^\mu) \mu_b(t) \, .
\label{Qb}
\end{equation}
The functions $p_b^\lambda$ and $p_b^\mu$ reflect a reduction in the efficiency of the feeding
(arrival) and the processing (departure) in production unit $b$. They
depend on $\rho_b$, $C_b$, $S_b$, and possibly other quantities as well. In some cases,
$p_b^\lambda$ is the probability of rejecting arrivals, 
when the storage capacity $S_b$ is reached, i.e. 
\begin{equation}
p_b^\lambda(\rho_b,C_b, S_b) = P_b(S_b) \,  , 
\end{equation}
where $P_b(l)$ is the probability of having a queue of length $l_b$ in production unit $b$. On the
other hand, the average number of active production channels is
\begin{equation}
  \sum_{l_b=0}^{C_b-1} l_b P_b(l_b) + \sum_{l_b=C_b}^{S_b} C_b P_b(l_b) \, .
\end{equation}
Consequently, the relative reduction $p_b^\mu$ in the processing rate due to $(C_b - l_b)$
empty channels is
\begin{equation}
 p_b^\mu(\rho_b,C_b, S_b) = \sum_{l_b=0}^{C_b-1} \frac{C_b - l_b}{C_b} \, P_b(l_b) \, .
\end{equation}
In more complex production systems, the factors $p_b^\lambda$ 
and $p_b^\mu$ would be generalized measures of inefficiencies 
in the production process, which would be functions of all relevant process parameters.
\par 
According to Little's Law $L_b = \overline{\lambda_b} T_b$ for queueing systems, 
where $\overline{\lambda_b} = (1-p_b^\lambda) \lambda_b$ 
is the average arrival rate, we can express the throughput also in terms of the average queue length
$L_b(\rho_b,C_b,S_b)$ and the average waiting time $T_b(\rho_b,C_b,S_b)$:
\begin{equation}
 Q_b(t) = \frac{L_b(\rho_b,C_b,S_b)}{T_b(\rho_b,C_b,S_b)} \, .
\label{Pb}
\end{equation}
For example, for a system with infinite storage capacity $S_b \rightarrow \infty$, we have
\begin{equation}
 Q_b = \min (\lambda_b, \mu_b) = \mu_b \min ( \rho_b, 1 ) \, .
\end{equation}
For a $M/M/1:(S_b/\mbox{FIFO})$ process  (one channel with first-in-first-out
serving, storage capacity $S_b$, Poisson-distributed arrival times and
exponentially distributed service intervals), one finds for $\lambda_b \le \mu_b$
\begin{equation}
 Q_b = \lambda_b \frac{1 - \rho_b{}^{S_b}}{1-\rho_b{}^{S_b+1}} \stackrel{\rho_b\rightarrow 1}{\longrightarrow}
 \lambda_b \frac{S_b}{S_b+1} \, . 
\end{equation}
Note that both, the expected value and the standard deviation of the 
queue length and the waiting time diverge for $\rho_b \rightarrow 1$. 
Therefore, efficient production is normally related to $\rho_b \le 0.7$ \cite{queuing}.

\subsection{The Feeding Rates of Production Units}

\begin{figure}[h]
\begin{center}
\includegraphics[width=10cm]{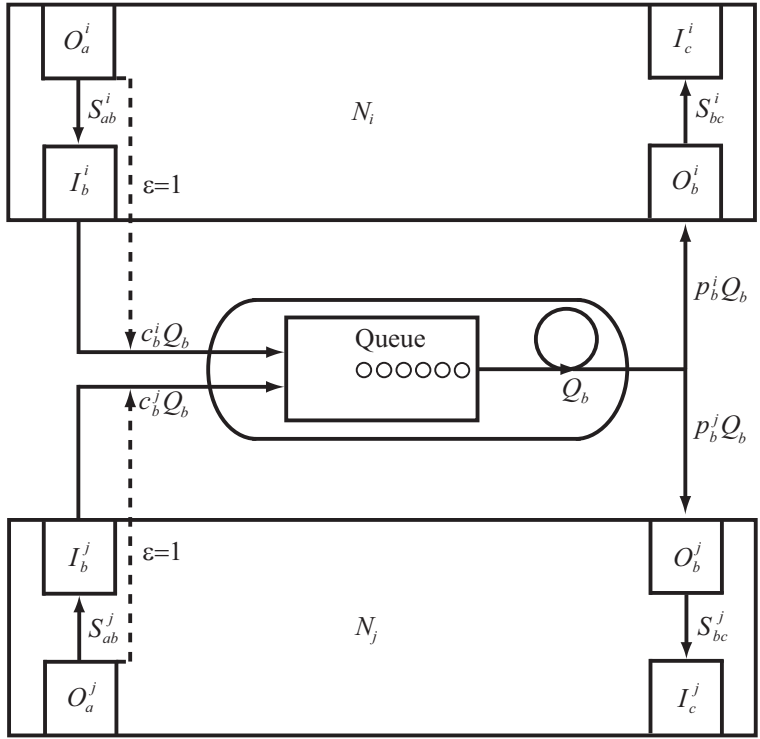} 
\end{center}
\caption{Schematic illustration of the product flows 
together with the model variables related to production unit $b$, which is represented by the
oval structure in the center. Rectangles correspond to buffers, arrows to
product flows.}
\label{figure1}
\end{figure}

In the absence of capacity constraints, we just have the relation $\lambda_b = \rho_b^0 \mu_b$ for the feeding
rate, where the actual utilization $\rho_b$ agrees with the desired utilization $\rho_b^0$,  
e.g. $\rho_b^0 = 0.7$. However, a lack of required educts may lead to a reduction of $\lambda_b$.
In that case, the feeding rate is limited by the minimum arrival rate $A_{b}^j$ of required educts $j$,
divided by the quantity $c_b^j$ of educts needed for one production cycle. 
We will assume that the maximum arrival rates $A_{b}^j$ are given by $\rho_b^0 \mu_b V_b^j I_b^j$,
where $I_b^j$ denotes the quantity of educts stored in the input buffer, 
$\rho_b^0 \mu_b V_b^j$ is the maximum transport rate due to capacity
contraints $V_b^j$ for getting educt $j$ from the input buffer into the production unit $b$,
and the prefactor $\rho_b^0 \mu_b$ suggests that the transport capacity is adapted to the production
speed (generalizations are possible). Therefore, if $\lambda_b^j(t) = \rho_b^0 \mu_b V_b^j I_b^j(t)/c_b^j$, 
denotes  the maximum feeding rate for educt $j$ into production unit $b$, the actually resulting feeding rate is 
\begin{equation}
 \lambda_b(t) = 
   \min ( \rho_b^0 \mu_b  , \{ \lambda_b^j \} ) \, ,
\end{equation}
where the minimum extends over all indices $j$. The product flows are illustrated in Fig.~\ref{figure1}.
\par
When we assume that the flows $S_{ab}^j$ of products $j$ to $b$ from the delivering production units $a$ can be
fed into the production process in parallel (instead of having to go through the input buffer first),
we have the generalized relationship 
\begin{equation}
 \lambda_b^j(t) = \left(\rho_b^0 \mu_b V_b^j I_b^j(t) + \varepsilon \sum_a S_{ab}^j(t) \right) \bigg/ c_b^j  
\end{equation}
with $\varepsilon = 1$. 
The previously discussed case (requiring delivery through the input buffer) corresponds to 
$\varepsilon = 0$. In any case, the fraction of the supply 
$\sum_a S_{ab}^j$, which is not needed to satisfy the production requirements
$c_b^j Q_b$, is delivered to the input buffer.

\subsection{Input and Output Buffers}

Let us assume each production unit $b$ has input buffers for required products $i$
(``educts'') and output buffers (a ``warehouse'')
for the products. We will assume the input buffers are filled with $I_b^i(t)$ educts
$i \in \{1,\dots,p\}$ and the output buffers with $O_b^i(t)$ products waiting to be delivered.
If $S_{ab}^i(t)$ denotes the delivery flow (``supply'')
of products $i$ from production unit $a$ to $b$, the change of
an input buffer stock with time is given by the conservation equation
\begin{equation}
  \frac{dI_b^i}{dt} = \sum_a S_{ab}^i(t) -  c_b^i Q_b(t)\, ,
\label{bal1}
\end{equation} 
as $\sum_a S_{ab}^i(t)$ is the quantity of products $i$ delivered from various sources (production units) $a$, 
and  $c_b^i Q_b$ is the quantity of educts $i$ used up for production per unit time (see Fig.~\ref{figure1}).
Analogously, the dynamics of an output buffer stock is determined by the equation
\begin{equation}
 \frac{dO_b^j}{dt} = p_b^j Q_b(t) - \sum_c S_{bc}^j(t) \, ,
\label{bal2}
\end{equation} 
as $p_b^j Q_b$ is the quantity of newly generated products $j$, and $\sum_c S_{bc}^j(t)$ are deliveries
to other production units $c$ (see Fig.~\ref{figure1}). 
The following specifications in this paper ensure the non-negativity conditions 
$I_b^i(t) \ge 0$ and $O_b^i(t) \ge 0$ (see also Secs.~\ref{SEC0} and
\ref{Sec2}). We will, for example, assume that the delivery flows $S_{ab}^i$  
are basically given by the order flows (``demands'') $D_{ab}^i$, but limited by the maximum transport rates
$V_{ab}^i\rho_b^0 \mu_b$ for delivering the available products from output buffer $O_a^i$ to production
unit $b$:
\begin{equation}
 S_{ab}^i(t) = \min [ D_{ab}^i(t), O_a^i(t) V_{ab}^i \rho_b^0 \mu_b] \, .
\label{Sab}
\end{equation}
If required, suitable specifications can also guarantee 
$I_b^i(t) \le I_b^{i,{\rm max}}$ and $O_b^i(t) \le O_b^{i,{\rm max}}$ with maximum
input and output buffer stocks $I_b^{i,{\rm max}}$ and $O_b^{i,{\rm max}}$. We may, for example,
multiply $S_{ab}^i$ by $[1 - (I_b^i/I_b^{i,{\rm max}})^\kappa ]$,
which stops the delivery flow of product $i$ when the respective input buffer of production unit $b$ is full.
Analogously, we may multiply $\rho_b^0\mu_b$ by $\prod_j [1 - (O_b^j/
O_b^{j,{\rm max}})^\kappa ]$,
which stops the production when one of the output buffers of $b$ is full. High values of $\kappa$
produce a hard cutoff, while small values of $\kappa \ge 1$ 
can describe cases, where the production efficiency goes down even
before the buffer size is fully used up. 

\subsection{Adaptation of Production Speeds and Transport Capacities}

One important aspect is the adaptation of production to changing demand. On the one hand,
a change of the processing rate $\mu_b$ is expensive and time consuming. On the other hand, the adaptation of the 
desired utilization $\rho_b^0$ is delayed by the average queuing time $T_b$ and other factors. 
We will assume 
\begin{equation}
 \frac{d(\rho_b^0\mu_b)}{dt} = \frac{1}{\tau_b} \left[ 
 W_b\!\left( 1/Z_b, \dots \right) - \rho_b^0 \mu_b \right] 
\label{zwei}
\end{equation}
with a typical adaptation time $\tau_b$ 
significantly greater than the adaptation times of the other variables. (That is, why we do not consider possible
time delays in the other mathematical relations, here.)
$W_b(1/Z_b,\dots)$ is some control function reflecting the strategy by the production manager
in adapting the utilization and/or processing rate to the output buffer stocks $O_a^j$ 
or other variables. In the following, we will assume 
\begin{equation}
\frac{1}{Z_b} = \sum_j \frac{p_b^j X_b^j}{N_b^{j}} \, ,
\end{equation}
i.e. the management strategy is only sensitive to the perceived stock levels $N_b^{j}$ of the products
$j$ generated by unit $b$, for which $p_b^j > 0$.
$X_b^j$ are proportionality constants.
The perceived stock levels will be specified by
\begin{equation}
  N_b^{j} = O_b^j + \delta \sum_{a(\ne b)} O_a^j \, . 
\end{equation}
For $\delta = 0$, the management takes into account only the own output buffer stock $O_b^j$, while
for $\delta = 1$, it tracks also the output buffer stocks  $O_a^j$ of competing production units $a$.
Normally, the control function $W_b$ will increase with decreasing perceived stock levels
$N_b^{j}$, but the function $W_b(1/Z_b,\dots)$ saturates
due to financial, spatial, or technological limitations and inefficiencies in the processing of high order flows.
Again, we will use a function of the form
\begin{equation}
W_b(1/Z_b,\dots) = \max\left(A_b \frac{1+B_b Z_b}{1+B_b Z_b+D_b (Z_b)^2}, 0\right)
\end{equation}
with suitable parameters $A_b$, $B_b$, and $D_b$. $A_b$ corresponds to the maximum
production speed. 

\subsection{Order Flows, Delivery Networks, and Price Dynamics} \label{SEC0}

The flow of orders is basically given by the flow $c_b^i Q_b$ of educts
required for the production by unit $b$. Deviations can be reflected by a
correction function $U_b(I_b^i,I_b^{i0})$, where $I_b^{i0}$ denotes the desired input
buffer stock. If the proportions $q_{ab}^i$ with $\sum_a q_{ab}^i = 1$ reflect orders 
from different producers $a$, we have
\begin{equation}
D_{ab}^i = q_{ab}^i c_b^i Q_b U_b(I_b^i,I_b^{i0})
\end{equation}
and, therefore,  
\begin{equation}
 S_{ab}^i(t) = \min [ q_{ab}^i c_b^i Q_b U_b(I_b^i,I_b^{i0}), O_a^i V_{ab}^i \rho_b^0 \mu_b ] 
\end{equation}
(see Eq.~(\ref{Sab})). When the first term is the smaller one, this implies with (\ref{bal1})
\begin{eqnarray}
\qquad \qquad \frac{dI_b^i}{dt} &=& \sum_a q_{ab}^i c_b^i Q_b U_b(I_b^i,I_b^{i0}) - c_b^iQ_b \nonumber \\
 &=& [ U_b(I_b^i,I_b^{i0}) - 1] c_b^i Q_b(t) \, . 
\label{Inp}
\end{eqnarray}
The function $U_b(I_b^i,I_b^{i0})$ should be chosen greater than 1, if $I_b^i$ is smaller than the
desired input buffer stock $I_b^{i0}$, smaller than 1 for $I_b^i> I_b^{i0}$, and otherwise 1, e.g. 
\begin{equation}
U_b(I_b^i,I_b^{i0}) = I_b^{i0}/I_b^i \, . 
\end{equation}
Without this correction function, the input buffer tends to be emptied in the course of time.
\par
Together with $c_b^i$ and $p_b^i$,
the fractions $ q_{ab}^i$ characterize the delivery or supply network. One can imagine
various scenarios. For example, the fractions could be assumed constant or
modeled by an evolutionary selection equation with a selection rate $\nu$ \cite{evol}: 
\begin{equation}
 \frac{dq_{ab}^i}{dt} = \nu \left( F_{ab}^i(t) - \sum_{a'}
  F_{a'b}^i(t) q_{a'b}^i(t) \right) q_{ab}^i(t) \, .
\end{equation}
The specification of the fitness depends on the relevant parameters. For example, it could be
treated constant.  However, in some cases,
it makes sense to relate the fitness $F_{ab}^i$ to the inverse of the real or virtual costs (``prices'')
$p_{ab}^i$ of product $i$, when delivered from production unit $a$ to $b$:
\begin{equation}
  F_{ab}^i(t) = 1/p_{ab}^i(t) \, .
\end{equation}
Assuming a law of supply and demand, one conceivable specification would be
\begin{equation}
 p_{ab}^i(t) = p_i^{0}  \, \frac{D_{ab}^i(t)}{S_{ab}^i(t)} 
= p_i^{0}  \, \frac{D_{ab}^i(t)}{\min [ D_{ab}^i(t), O_a^i(t) V_{ab}^i(t) \rho_b^0 \mu_b ]} \, , 
\end{equation}
where $p_i^0$ are the costs when the supply $S_{ab}^i$ agrees with the demand $D_{ab}^i$. 
For example, a reduction in the price level would require a slight generalization of Eq.~(\ref{Sab}).
\par
Other specifications are, of course, possible as well,
as the above formulas partly depend on the strategies of the human decision makers involved. 

\subsection{Calculation of the Cycle Times}

Apart from the productivity or throughput $Q_b$ of a production unit, production managers are
also highly interested in the cycle time, i.e. the time interval between the beginning of the generation
of a product and its completion. Let us first discuss the process cycle time $t_b$ between entering the queue of
production unit $b$ and leaving it, assuming that all $c_b^i$ required educts $i$ for one production cycle
are transported together and located at the same place in the queue. The problem is similar to
determining the travel times of vehicles entering a traffic jam.
\par
According to Eq.~(\ref{Pb}), the average waiting time is determined as the quotient $T_b = L_b/Q_b$
of the average queue length $L_b$ and the
throughput $Q_b$, if production operates in the steady state. Let us now generalize this formula 
to situations in which the inflow
\begin{equation}
 Q_b^{\rm in} = (1-p_b^\lambda) \lambda_b
\end{equation}
into production unit $b$ is time-dependent and possibly differs from the time-dependent outflow
\begin{equation}
 Q_b^{\rm out} = (1-p_b^\mu) \mu_b 
\end{equation}
(see Fig.~\ref{ADD}). In reality, this time-dependence results from fluctuations in the production process and 
breakdowns of machines, etc.  In some cases, one can use the length-dependent formulas
\begin{equation}
 Q_b^{\rm in}(l_b) = \left\{
\begin{array}{ll}
\lambda_b & \mbox{if } l_b < S_b \\
0 & \mbox{otherwise}
\end{array} \right.
\end{equation}
and
\begin{equation}
 Q_b^{\rm out}(l_b) = \left\{
\begin{array}{ll}
\mu_b & \mbox{if } l_b \ge C_b \\
\mu_b \, l_b/C_b & \mbox{if } l_b < C_b 
\end{array} \right.
\end{equation}
(see Sec.~\ref{SEC2a}).
We will now derive a delay-differential equation for the cycle time under varying production conditions. 
For this, let $l_b(t)$ be the actual length of the queue in production unit $b$ at time $t$. The change of this
length in time is given by the difference between the inflow and the outflow
at time $t$:
\begin{equation}
 \frac{dl_b}{dt} = Q_b^{\rm in}(t) - Q_b^{\rm out}(t) \, .
\label{length}
\end{equation}
If the production
unit $b$ has $C_b$ channels, an educt entering the production queue at time $t$ 
must move forward $l_b(t)-C_b$ steps, before it is finally served
by one of the $C_b$ channels. If, after entering the queue at time $t$, 
$t^*_b(t)$ denotes the waiting time until one of the channels is reached,
and if the average processing rate of a single channel is $\mu_b/C_b$, 
the average serving or treatment time is given by $C_b/\mu_b(t+t^*_b)$. The overall time $t_b$ required for the
processing of the product is, therefore, given by the sum of the waiting time $t^*_b$ and the treatment time
by one of the channels:
\begin{equation}
 t_b(t) = t^*_b(t) + \frac{C_b}{\mu_b(t+t^*_b(t))} 
\end{equation}
(which replaces the average value $T_b$).
On the other hand, the waiting educts move forward $Q_b^{\rm out}$ steps per unit time. For this reason,
the waiting time $t^*_b(t)$ is given by the implicit equation
\begin{equation}
 l_b(t) - C_b = \int\limits_t^{t+t^*_b(t)} \!\! dt' \, Q_b^{\rm out}(t') 
=  \int\limits_{-\infty}^{t+t^*_b(t)} \!\! dt' \, Q_b^{\rm out}(t') 
 - \int\limits_{-\infty}^{t} \!\! dt' \, Q_b^{\rm out}(t') \, .
\end{equation}
Identifying the time-derivative of this equation with Eq.~(\ref{length}) results in 
\begin{equation}
 Q_b^{\rm in}(t) - Q_b^{\rm out}(t) = \frac{dl_b}{dt} = Q_b^{\rm out}(t+t^*_b(t))\left(1+\frac{dt^*_b}{dt}\right)
 - Q_b^{\rm out}(t) \, ,
\end{equation} 
which leads to the delay-differential equation
\begin{equation}
 \frac{dt^*_b}{dt} = \frac{Q_b^{\rm in}(t)}{Q_b^{\rm out}(t+t^*_b(t))} - 1 \, .
\end{equation}
As the production initially starts with a waiting time of $t^*_b(0) = 0$ (when the factory or production
unit $b$ is opened), this equation can be solved numerically as a function of the outflow $Q_b^{\rm out}(t')$.
In this way, it is possible to determine the waiting time $t^*_b(t)$ and process cycle time $t_b$. In the future, 
approximation methods shall be developed for cases where $t^*_b(0)$ is not known or $Q_b^{\rm out}(t')$ cannot
be controlled or anticipated. A rough approximation would be to replace these values by mean values
(cf. Eq.~(\ref{Pb})).
\par
In a similar way, one can calculate the waiting time $t_b^i$ in the input buffer $I_b^i$, resulting in
\begin{equation}
 \frac{dt_b^i}{dt} = \frac{\sum_a S_{ab}^i(t)}{c_b^iQ_b^{\rm out}(t+t_b^i(t))} - 1 \, .
\end{equation}
The waiting time $t_b^j$ in the output buffer $O_b^j$ is given by
\begin{equation}
 \frac{dt_b^j}{dt} = \frac{p_b^j Q_b(t)}{\sum_c S_{bc}^j (t+t_b^j(t))} - 1 \, .
\end{equation}
Finally, the transport time $t_{ab}^i$between the output buffer of production unit $a$ and the input buffer of
production unit $b$ can be estimated by
\begin{equation}
  t_{ab}^i = \frac{1}{V_{ab}^i} \, .
\end{equation}
The total cycle time is, then, calculated as the sum of the respective waiting, serving, and transport times.

\section{Relation between Production Processes and Macroeconomics} \label{SEC3}

We will now give some support for the ``macroscopic'' model of supply networks applied in Sec.~\ref{Sec1}
by relating it to the ``microscopic'' dynamical production model developed in Sec.~\ref{SEC2}.

\subsection{Just-in-Time Production} \label{Sec2}

Let us start with deriving a simplified model of production processes. 
For this, we will summarize the input and output buffers,
defining the stock levels (``inventories'') of markets for the products $i$ by 
\begin{equation}
 N_i(t) = \sum_b I_b^i(t) + \sum_b O_b^i(t) \, .
\label{micmac}
\end{equation}
According to the balance equations (\ref{bal1}) and (\ref{bal2}), we find
\begin{equation}
 \frac{dN_i}{dt} = \sum_b (p_b^i - c_b^i) Q_b(t) \, .
\label{mac}
\end{equation}
Moreover, typical for just-in-time production, we will assume negligible input buffers
\begin{equation}
 I_b^i = I_b^{i0} = 0, \quad \mbox{which implies} \quad N_i = \sum_b O_b^i \, .
\end{equation}
Consequently, for $\varepsilon = 1$ and $V_{ab}^i = V_b^i$, the feeding (arrival) rates become
\begin{equation}
 \lambda_b(t) =   \rho_b^0 \mu_b \min \left( 1 , \left\{ \frac{V_b^j N_j(t)}{c_b^j}\right\} \right) \, .
\label{merk} 
\end{equation}
Furthermore, for $\delta = 1$ we obtain $N_b^j = \sum_a O_a^j = N_j$ and the related adaptation equations
\begin{equation}
 \frac{d(\rho_b^0\mu_b)}{dt} = \frac{1}{\tau_b} \left[ 
 W_b\!\left( \sum_j \frac{p_b^j X_b^j}{N_j(t)} ,
 \dots \right) - \rho_b^0 \mu_b \right] \, .
\label{eq1}
\end{equation}
\par
We will now assume $Q_b = \lambda_b$, which is usually given for
small enough channel utilizations $\rho_b$ or sufficient storage capacities $S_b$. 
In that case, one would have to solve Eq.~(\ref{eq1}) together with
\begin{equation}
 \frac{dN_i}{dt} = \sum_b (p_b^i - c_b^i) \rho_b^0 \mu_b \min 
\left( 1 , \left\{\frac{V_b^j N_j(t)}{c_b^j} \right\}\right) \, .
\label{formula}
\end{equation}
Note that formula (\ref{formula}) maintains
non-negative inventories $N_i(t)$, as required. To show this, let us assume that $t$ would be the first
point in time where some inventory $N_j$ vanishes, say $N_i(t) = 0$. The inventory $N_i(t)$ could
only become negative for $dN_i/dt < 0$, which would require $c_b^i > 0$. But then, the minimum
in formula (\ref{formula}) would be $V_b^i N_i(t)/c_b^i = 0$, which contradicts our assumption and
proves non-negativity (at least for $V_b^j > 0$). 

\subsection{Micro-Macro Link}

In the following, we will try to reduce the number of equations by an aggregation procedure, 
which keeps the structure of the above model equations. Let us first define production sectors $B$
by summarizing those production units $b$, which are characterized by the same adaptation times
$\tau_b$ and a proportional throughput $Q_b(t)$. With suitable constants
$\tau_B$ and $k_b$, we can then assume
\begin{equation}
 \tau_b = \tau_B \quad \mbox{and} \quad Q_b(t) = k_b Q_B(t) \, . 
\label{ass1}
\end{equation}
The total throughput, processing and feeding rates of production sector $B$ are defined by 
\begin{equation}
 Q_B = \sum_{b\in B} Q_b \, , \quad \mu_B = \sum_{b\in B} \mu_b \, , 
\quad \lambda_B = \sum_{b\in B} \lambda_b \, ,
\end{equation}
where $b\in B$ indicates that $b$ belongs to production sector $B$. The equations for the
inventories keep their structure
\begin{equation}
\frac{dN_i}{dt} = \sum_B (p_B^i - c_B^i) Q_B(t) \, ,
\label{I}
\end{equation}
if we define
\begin{equation}
 p_B^i = \sum_{b\in B} p_b^i k_b\quad \mbox{and} \quad c_B^i = \sum_{b\in B} c_b^i k_b 
\end{equation}
with $k_b = Q_b(0)/Q_B(0)$ and use the relation $\sum_b = \sum_B \sum_{b\in B}$.
Moreover, in order to get the equations
\begin{equation}
 \lambda_B(t) =    \rho_B^0 \mu_B \min \left( 1, \left\{ \frac{V_B^j N_j(t)}{c_B^j} \right\}\right) 
\end{equation}
and
\begin{equation}
 \frac{d(\rho_B^0\mu_B)}{dt} = \frac{1}{\tau_B} \left[ 
 W_B\!\left( \sum_j \frac{p_B^j X_B^j}{N_j(t)} ,
 \dots \right) - \rho_B^0 \mu_B \right] \, ,
\end{equation}
we need to set 
\begin{equation}
 V_B^j = V_b^j c_B^j / c_b^j  \quad \mbox{and} \quad X_B^j = X_b^j p_b^j/p_B^j  
\label{satis}
\end{equation}
for all $j$ (requiring that the resulting values on the right-hand sides depend only on $B$, but not on $b\in B$), and 
\begin{equation}
\rho_B^0 = \sum_{b\in B} \rho_b^0 \mu_b/\mu_B \, , \quad W_B(1/Z_B) = \sum_{b\in B} W_b(1/Z_b) 
\end{equation}
with $1/Z_B = \sum_j p_B^j X_B^j/N_j = \sum_j p_b^j X_b^j/N_j = 1/Z_b$.
\par
We may also summarize all those markets $i$ which show a proportional dynamics of
$N_i(t)$ in time. This assumes 
\begin{equation}
  N_i(t) = X_i N_I(t) \, ,
\label{ass2}
\end{equation}
if we denote the proportionality factors by $X_i$ and define
\begin{equation}
 N_I(t) = \sum_{i\in I} N_i(t) \, , 
\end{equation}
where $i\in I$ indicates that $i$ is part of the market sector $I$. Introducing
\begin{equation}
 p_B^I = \sum_{i\in I} p_B^i \quad \mbox{and} \quad c_B^I = \sum_{i\in I} c_B^i 
\end{equation}
and summing up Eq.~(\ref{I}) over ${i \in I}$ yields the further reduced set of equations
\begin{equation}
\frac{dN_I}{dt} = \sum_B (p_B^I - c_B^I) Q_B(t) \, .
\label{BE}
\end{equation}
In order to obtain the equations
\begin{equation}
 \lambda_B(t) =    \rho_B^0 \mu_B \min \left( 1, \left\{ \frac{V_B^J N_J(t)}{c_B^J} \right\}\right) 
\end{equation}
and
\begin{equation}
 \frac{d(\rho_B^0\mu_B)}{dt} = \frac{1}{\tau_B} \left[ 
 W_B\!\left( \sum_J \frac{p_B^J X_B^J}{N_J(t)} ,
 \dots \right) - \rho_B^0 \mu_B \right] \, ,
\end{equation}
we have to define
\begin{equation}
 V_B^J = \min_{j\in J} \left\{ \frac{V_B^j c_B^J X_j}{c_B^j} \right\}
 = \min_{j\in J} \left\{ \frac{V_b^j c_B^J X_j}{c_b^j} \right\}
\end{equation}
and
\begin{equation}
 X_B^J = \sum_{j\in J} \frac{p_B^j X_B^j}{p_B^J X_j} = \sum_{j\in J} \frac{p_b^j X_b^j}{p_B^J X_j} \, .
\end{equation}
Based on assumptions (\ref{ass1}), (\ref{satis}), and (\ref{ass2}), 
one could achieve a considerable reduction in the number of equations, leading from
a microscopic model of production processes involving management decisions
to a macroscopic model of interacting economic sectors. 
Conditions (\ref{ass1}), (\ref{satis}), and (\ref{ass2}) presuppose similar production processes of
the summarized production units $b$ and proportional coefficients regarding the summarized markets $i$.
While our aggregation method can be generalized to time-dependent parameters $k_b(t)$ and $X_i(t)$,
we still require $\tau_b = \tau_B$, and the resulting values of $V_b^j c_B^j/c_b^j$,
$X_b^j p_b^j/p_B^j$ should only depend on $B$, but not $b\in B$. These are the main criteria for defining
production and market sectors based on empirical data.

\subsection{Dynamic Input-Output Model} \label{Sec3}

In Eq. (\ref{BE}), the sum over $B$ extends from 1 to $U+1$, with $B=U+1$ representing the
final consumer sector. Splitting it up, the resulting equations read
\begin{equation}
 \frac{dN_I}{dt} = \sum_{B=1}^U (p_B^I - c_B^I) Q_B(t) - c_{U+1}^I Q_{U+1}(t) \, ,
\label{A}
\end{equation}
as $p_{U+1}^I = 0$. Similarly, $J$ in formula (\ref{merk})
runs from 0 to $P$ with $J=0$ representing a market sector for basic resources. Therefore,
\begin{equation}
 \lambda_B(t) = \rho_B^0(t) \mu_B  \min \left( 1 , \frac{V_{B}^0 N_0(t)}{c_{B}^0} ,
 \left\{\frac{V_B^J N_J(t)}{c_B^J} \right\} \right) \ge 0 \, ,
\label{mini}
\end{equation}
where $J\in \{1,\dots,P\}$. 
\par
\begin{figure}[h]
\begin{center}
\includegraphics[width=10cm]{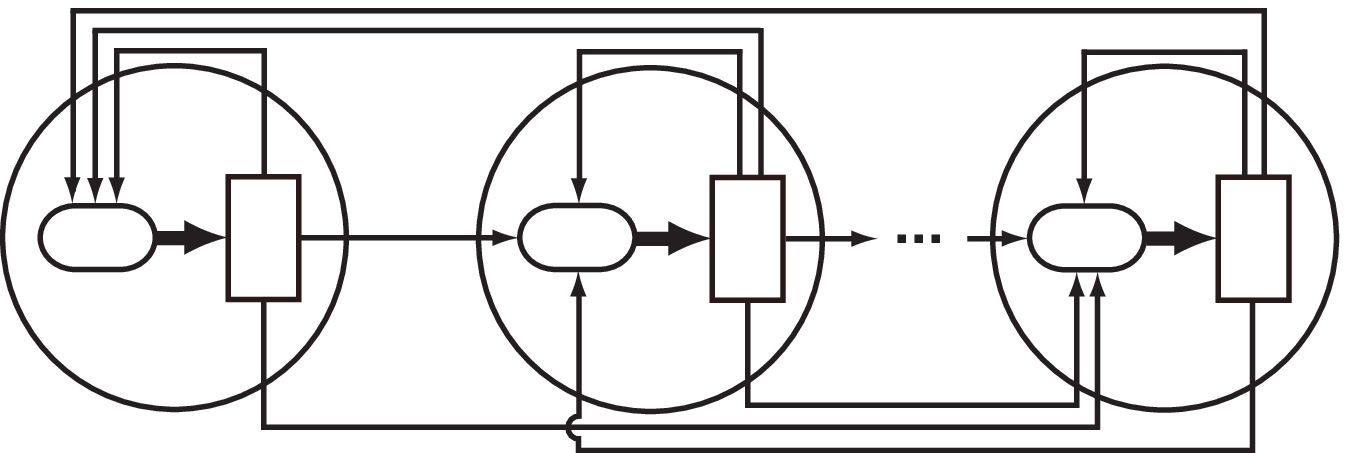}\\[4mm] 
\includegraphics[width=10cm]{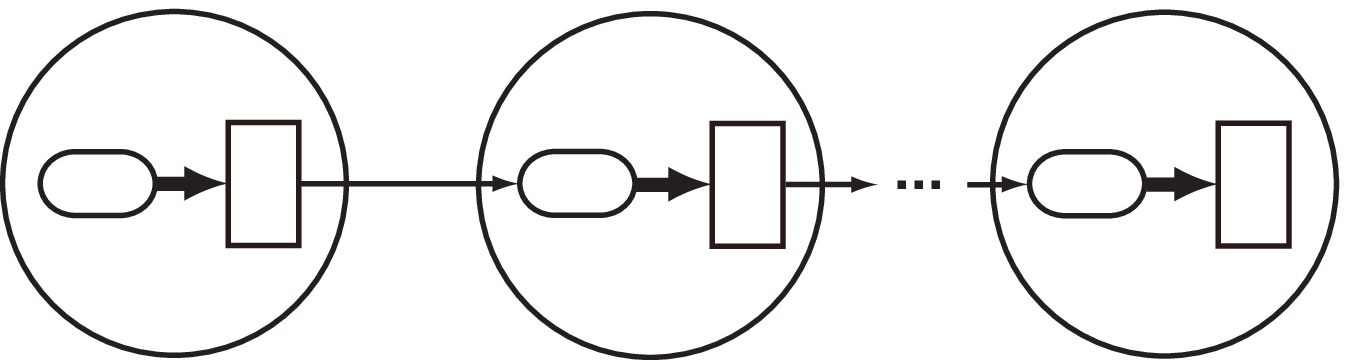} 
\end{center}
\caption{Top: Schematic illustration of the actual product flows according the the
input-output model with $p_B^I = \delta_{B,I}$. The circle summarizes the production sector (oval)
and its market (rectangle). The associated product flow 
is represented by a thick arrow, the product flows into the production sector by thin arrows. Bottom: Particular case
of a linear supply chain, in which each production unit receives goods only from the previous
one.}
\label{figure11}
\end{figure}

Let us now concentrate again on the case $Q_{B}(t) = \lambda_{B}(t)$
and define the consumption rate from market sector $I$ by
\begin{equation}
 Y_I(t) = c_{U+1}^I Q_{U+1}(t) \, .
\end{equation}
When we define the $U$ production sectors $B$ through the respective kind of products they produce,
we can set $p_B^I = k_B \delta_{B,I}$ (see Fig.~\ref{figure11}). 
Without restriction of generality, we can choose $k_B = 1$, which
just defines the unit quantity in which we measure products of market sector $I$.
In the stationary case with $dN_I/dt=0$ for all $I$, we then obtain the equation
\begin{equation}
  \sum_{B=1}^U c_B^I Q_B + Y_I = \sum_{B=1}^U p_B^I Q_B =  Q_I \, ,
\label{Leo}
\end{equation}
which corresponds to Leontief's 
input-output model of macroeconomics \cite{Leontief}. Therefore, the above model
of supply networks can be considered as dynamical generalization of this classical economic model.
\par
We should note that, in view of the instability of supply networks, i.e. the existence of the bullwhip 
effect and of business cycles, the applicability of steady-state concepts in economics is questionable.
People have, therefore, tried to formulate dynamical input-output models for a long time in order to
take into account investment strategies and other aspects. However, many of these approaches have
turned out to be inconsistent or useless (cf. Ref.~\cite{InOut}). 
In contrast, the above dynamical input-output model results
naturally from a much more general model of supply and production networks. Investment strategies
could, for example, be taken into account by a suitable specification
of $W_B(\dots)$, which may not only be
chosen as a function of the inventories $N_I$, but also 
of time derivatives such as $dQ_I/dt$, $dY_I/dt$, or $dN_I/dt$. 

\subsection{Specification of the Boundary Conditions} \label{SEC3d}

In the following, we will discuss some examples.
If we assume something like a conservation of materials or value, this implies certain constraints,
which reduce the potential complexity of the related supply networks. 
First of all, the quantity of products $I$ consumed by the production sectors $B$ 
should be generated somewhere, i.e.
\begin{equation}
\sum_B c_B^I = \sum_B p_B^I \, . 
\label{ZWO}
\end{equation} 
Second, the quantity of educts consumed by some production sector $B$ corresponds to the
quantity of its generated products, i.e.
\begin{equation}
 \sum_I c_B^I = \sum_I p_B^I \, .
\label{ONE}
\end{equation}
For closed systems (with $c_B^0 = 0$ and $c_{U+1}^I = 0$), 
this implies that the sum over all inventories $N_I$ is constant, i.e.
\begin{equation}
 \sum_I \frac{dN_I}{dt} = \sum_B  Q_B(t) \sum_I (p_B^I - c_B^I) = 0 \, .
\end{equation}
In the following, we will again discuss the case $p_B^I = \delta_{B,I}$. 
In this particular case, relations (\ref{ONE}) and (\ref{ZWO}) imply the normalization conditions
\begin{equation}
  \sum_B c_B^I = 1 = \sum_I c_B^I \, .
\end{equation}
As in Sec.~\ref{Sec3}, we will extend the sums over 
$I$ from 0 to $U$ and the sums over $B$ from 1 to $U+1$. In this way, we define 
\begin{equation}
 c_B^0 = 1 - \sum_I c_B^I \quad \mbox{and} \quad
 c_{U+1}^I = 1 - \sum_B c_B^I \, .
\label{bound}
\end{equation}
Values $c_B^0> 0$ allow us to describe
the inflow of basic resources $I=0$, while
$c_{U+1}^I > 0$ allows one to describe the depletion of products by a consumer sector $B=U+1$,
the aging of products, or the generation of products of minor quality.
The boundary conditions are completely defined by specifying $N_0(t)$ and $Q_{U+1}(t)$, see Eqs.~(\ref{A})
and (\ref{mini}). This treatment is fully consistent with the intuitive one of linear supply chains in Sec.~\ref{bull}.

\section{Impact of the Supply Network's Topology}\label{Topo}

\begin{figure}
\begin{center}
\begin{picture}(6,12.5)(0,5)
\put(0.05,17){\includegraphics[width=5.8\unitlength,clip=]{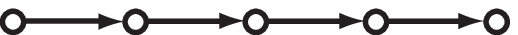}} 
\put(0.05,15){\includegraphics[width=5.8\unitlength,clip=]{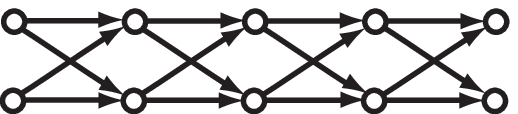}} 
\put(0.05,5.6){\includegraphics[width=5.8\unitlength,clip=]{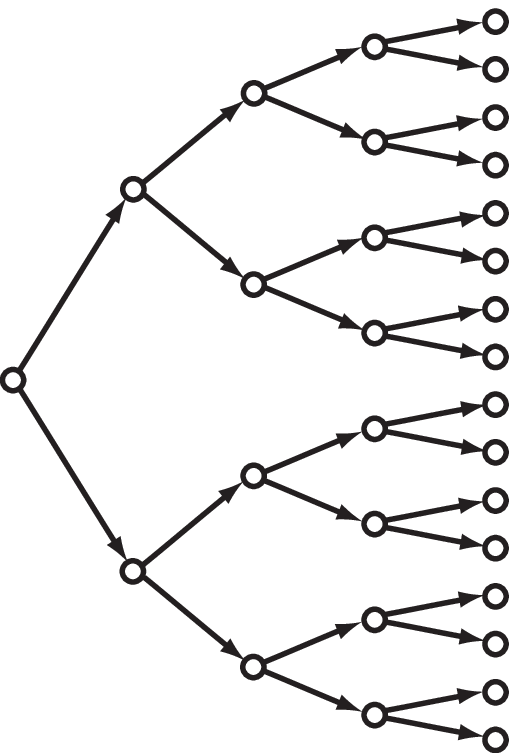}}
\put(0,17.5){{\footnotesize (a)}}
\put(0,16.5){{\footnotesize (b)}}
\put(0,12.1){{\footnotesize (c)}}
\end{picture}
\end{center}
\caption{Illustration of different supply networks: (a) linear supply chain with
five production units, (b) ``supply ladder'' with five levels, and (c) hierarchical supply tree. 
Circles summarize production sectors and their markets, as in Fig.~\ref{figure11}, while arrows
represent the product flows among the sectors.}
\label{Fig6}
\end{figure}

We will now discuss simulations based on the equations specified in Secs.~\ref{Sec2} to \ref{SEC3d}
for the three different supply networks sketched in Figs.~\ref{Fig6}a--c,
each with five levels: (a) a linear supply chain with 5 production units, 
(b) a ``supply ladder'' with 10 production units, and (c) a hierarchical supply
tree with 31 production units. 
\par\unitlength1cm
By introducting random variables $\xi_{K}^L$, which 
were assumed to be equally distributed in the interval $[-\eta,\eta]$, we have taken into
account a heterogeneity $\eta$ in the individual parameters characterizing the different production
units. Here, we have chosen $N_0(t) = N_0 = 20 = X_B^J$, $N_I(0) = 20 (1+\xi_I)$,
$\tau_B = 180 (1+\xi_B)$, $V_B^0 =  V = 0.05$, $V_B^J = V c_B^J (1+\xi_B^J)$ for $J>0$,  
$\rho_B^0(0) \mu_B = W_B(20/N_B(0))$, $Q_{U+1}(t) = W_{U+1}(20/N_0) \min(1,N_I(t) V /c_{U+1}^I) 
[1+0.1\sin(0.04 t)]$ and $A_B=100/V$. 
\par
\begin{figure}[h]
\begin{center}
\includegraphics[height=10cm,angle=-90]{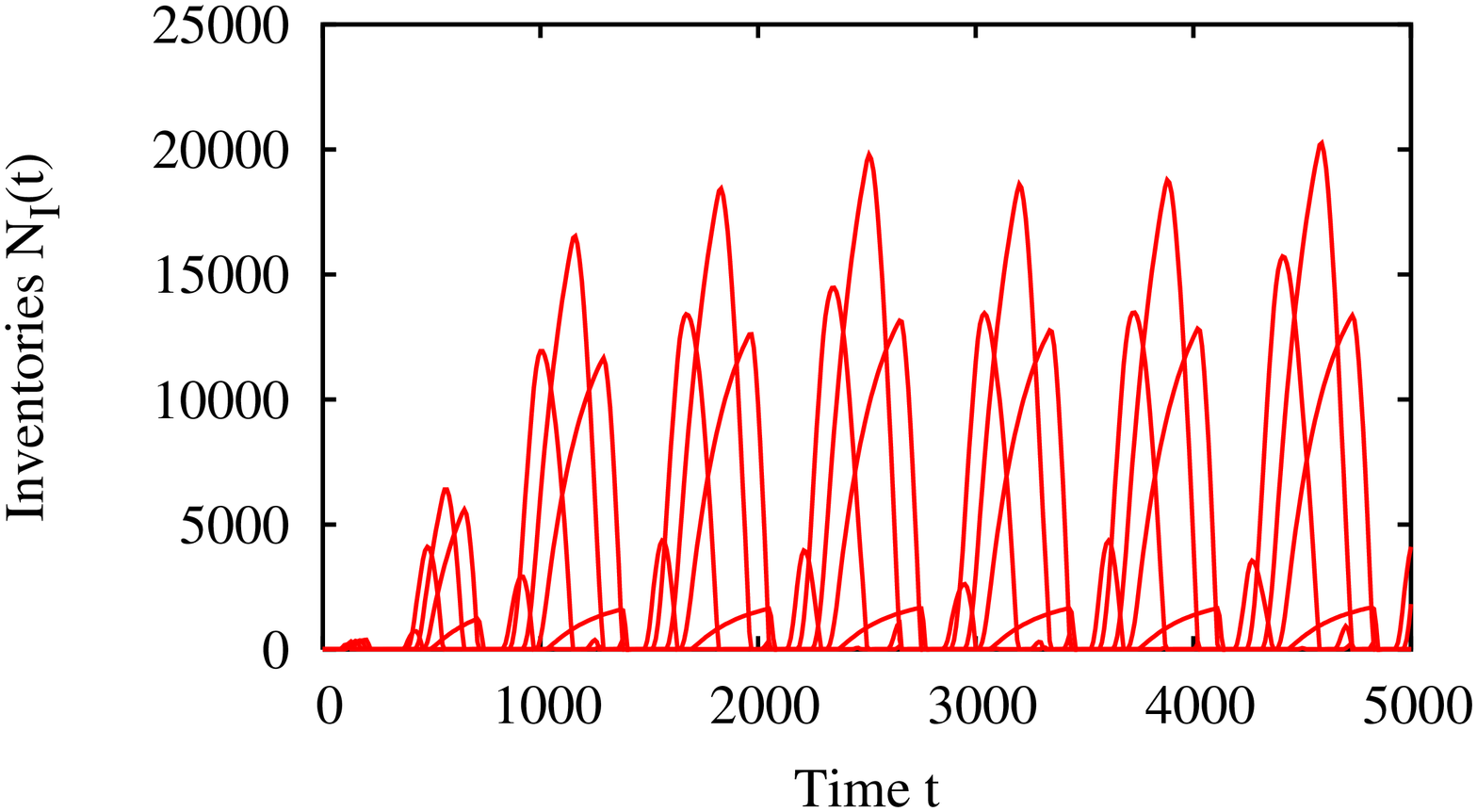} 
\end{center}
\caption{Time-dependent inventories for a supply network with five layers and 
identical model parameters ($\eta = 0$). The dynamics is the same for all three network topologies 
displayed in Fig.~\ref{Fig6}, i.e. for the linear supply chain, the supply ladder, and the hierarchical
supply network.} 
\label{Fig9}
\end{figure}
\begin{figure}[htbp]
\begin{center}
\includegraphics[height=9.8cm,angle=-90]{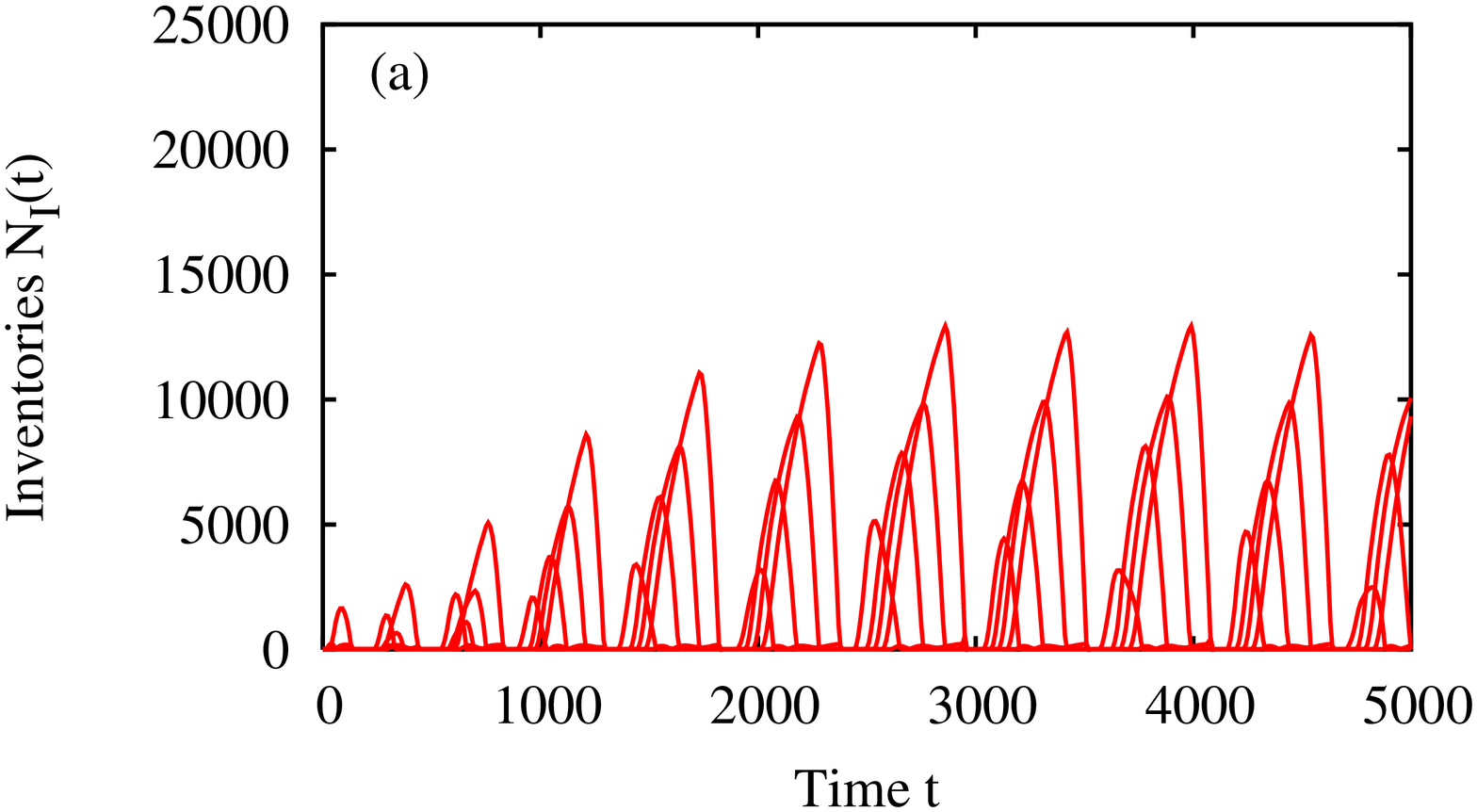}\\ 
\includegraphics[height=9.8cm,angle=-90]{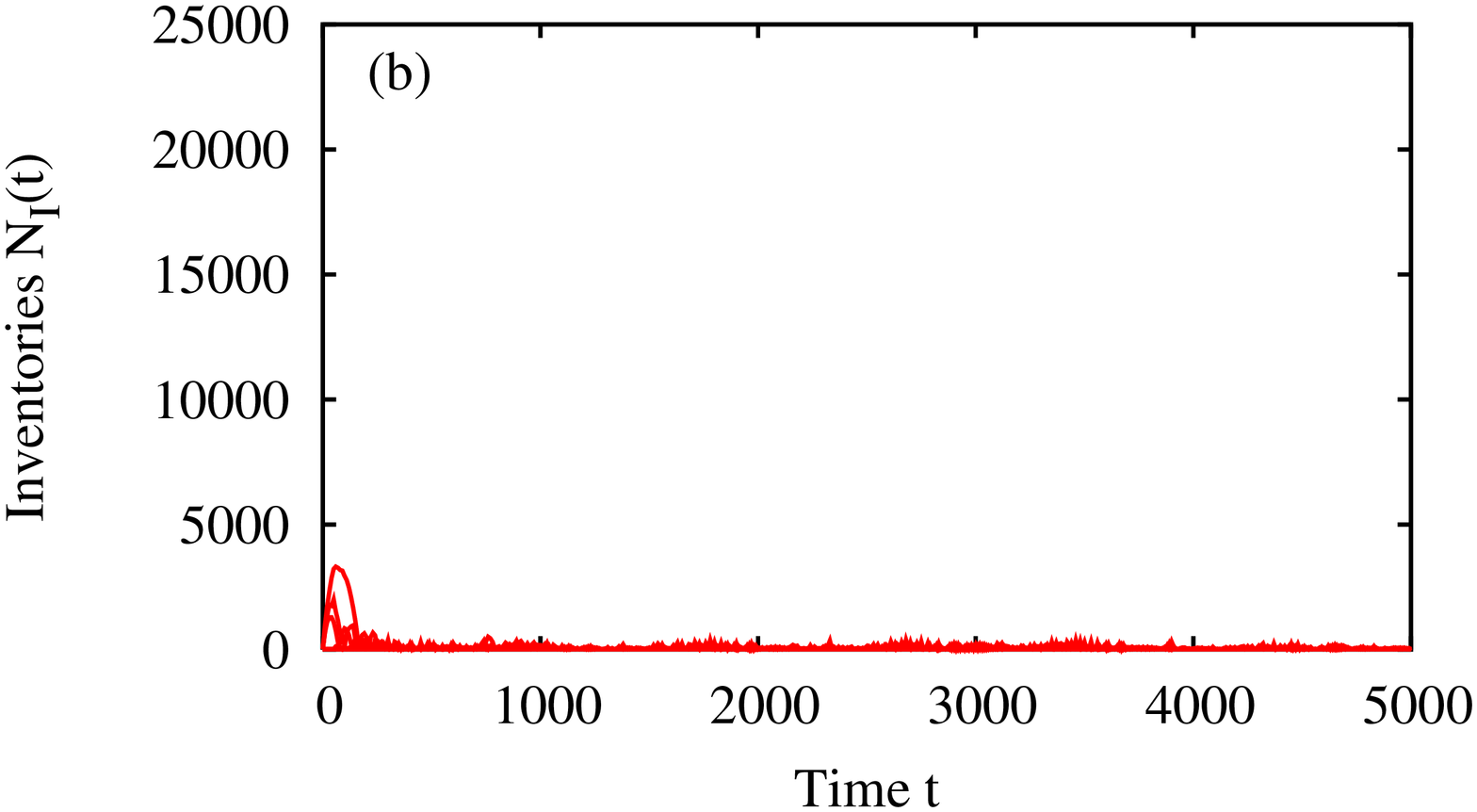}\\ 
\includegraphics[height=9.8cm,angle=-90]{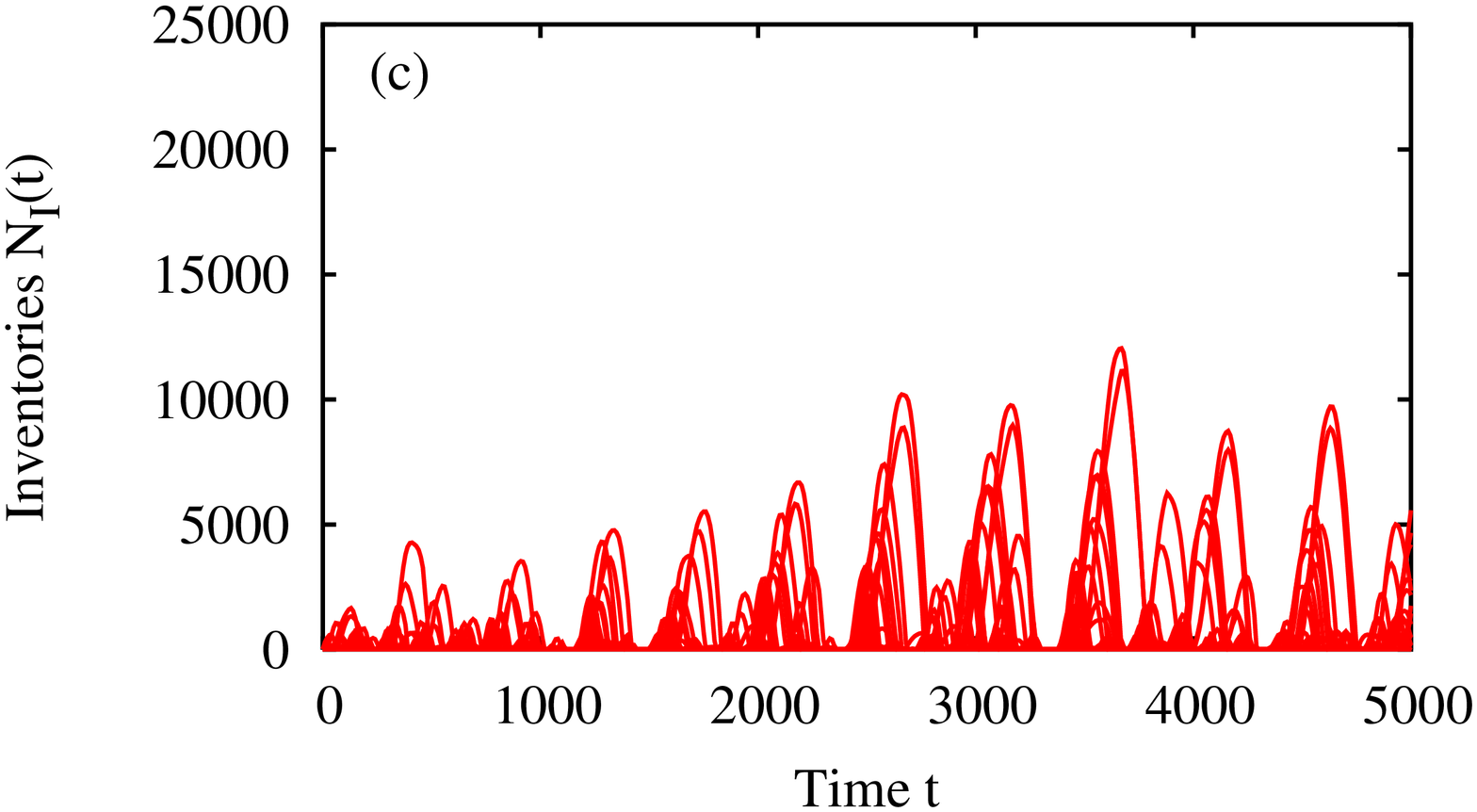} 
\end{center}
\caption{Time-dependent inventories for supply networks in the case of heterogeneous parameters
(with a heterogeneity of $\eta = 0.2$). The dynamics drastically depends on the topology of the
supply network: (a) linear supply chain, (b) supply ladder, (c) hierarchical supply network.
Compared with Fig.~\ref{Fig9} one can conclude that heterogeneity in supply networks can considerably decrease the
undesired oscillation amplitudes in the inventories. The strongest effect by far is found for supply ladders,
which is relevant for the design of robust supply networks.}
\label{Fig10}
\end{figure}

For the linear supply chain, we have $c_B^I=1$, if $I$ delivers to $b$, 
otherwise $c_B^I=0$:
\begin{equation}
{\mathbf{C}} = (C_{IB}) = (c_B^I) = \left(
\begin{array}{cccccc}
0.0 & 1.0 & 0.0 & 0.0 & 0.0 & 0.0 \\
0.0 & 0.0 & 1.0 & 0.0 & 0.0 & 0.0 \\
0.0 & 0.0 & 0.0 & 1.0 & 0.0 & 0.0 \\
0.0 & 0.0 & 0.0 & 0.0 & 1.0 & 0.0 \\
0.0 & 0.0 & 0.0 & 0.0 & 0.0 & 1.0 \\
0.0 & 0.0 & 0.0 & 0.0 & 0.0 & 0.0 \\ 
\end{array}
\right)
\end{equation}
For the supply ladder and
the hierarchical supply tree, we have $c_B^I = 0.5$, if an arrow points from 
$i$ to $b$ (see Figs.~\ref{Fig6}b, c), otherwise $c_B^I = 0$. Specifically, for the supply
ladder we have 
\begin{equation}
\mathbf{C} = \left(
\begin{array}{ccccccc}
0.0 & 0.0 & 0.5 & 0.5 & 0.0 & 0.0 & \dots \\
0.0 & 0.0 & 0.5 & 0.5 & 0.0 & 0.0 & \dots \\
0.0 & 0.0 & 0.0 & 0.0 & 0.5 & 0.5 & \dots \\
0.0 & 0.0 & 0.0 & 0.0 & 0.5 & 0.5 & \dots \\
0.0 & 0.0 & 0.0 & 0.0 & 0.0 & 0.0 & \dots \\
0.0 & 0.0 & 0.0 & 0.0 & 0.0 & 0.0 & \dots \\ 
\vdots & \vdots & \vdots & \vdots & \vdots & \vdots & \ddots \\
\end{array}
\right) \, ,
\end{equation}
while for the supply hierarchy, we have 
\begin{equation}
\mathbf{C} = \left(
\begin{array}{cccccccc}
0.0 & 0.5 & 0.5 & 0.0 & 0.0 & 0.0 & 0.0 & \dots \\
0.0 & 0.0 & 0.0 & 0.5 & 0.5 & 0.0 & 0.0 & \dots \\
0.0 & 0.0 & 0.0 & 0.0 & 0.0 & 0.5 & 0.5 & \dots \\
0.0 & 0.0 & 0.0 & 0.0 & 0.0 & 0.0 & 0.0 & \dots \\
0.0 & 0.0 & 0.0 & 0.0 & 0.0 & 0.0 & 0.0 & \dots \\
0.0 & 0.0 & 0.0 & 0.0 & 0.0 & 0.0 & 0.0 & \dots \\ 
0.0 & 0.0 & 0.0 & 0.0 & 0.0 & 0.0 & 0.0 & \dots \\ 
\vdots & \vdots & \vdots & \vdots & \vdots & \vdots & \vdots & \ddots \\
\end{array}
\right) \, .
\end{equation}
$c_B^0$ and $c_{U+1}^I$ 
are defined in accordance with Eq.~(\ref{bound}). These specifications
guarantee that, {\rm for $\eta = 0$,} i.e. if 
the production units are characterized by identical parameters, {\em the dynamics of the
inventories is the same for all three discussed network topologies} (see Fig.~\ref{Fig9}).
However, the topology 
matters a lot, if we have a heterogeneity $\eta > 0$ in the model parameters  (see Figs.~\ref{Fig10}e--f). 
As our dynamical model of supply networks assumes non-linear interactions, 
changes of $\eta$ can have large effects. The same
applies to small changes in $N_0$ (see Fig.~\ref{FIG8}) or
in the relaxation times $\tau_B$ (see Fig.~\ref{FIG3}).

\section{Summary and Outlook} \label{Sec4}

In this contribution, we have sketched a dynamical theory of supply networks. Here, we could only present
first results and indicate possible future research directions, which may contribute to the interdisciplinary field of
econophysics \cite{econophysics}. The proposed theory is developed to help understand the complex non-linear phenomena
in production and supply networks, in particular their breakdowns, instabilities and inefficiencies.
We have also sketched how to derive ``macroscopic'' equations for the dynamics of a sectorally structured
economics from ``microscopic'' equations describing single production processes, involving 
strategical decisions of production managers reflected by the control functions $W_b$. 
The resulting model is a dynamical generalization of the classical input-output model of macroeconomics.
For the particular case of a linear supply chain, one can relate it with
the Hilliges-Weidlich model, which has originally been developed for traffic flow. These equations can
describe the ``bullwhip effect'' due to their linear instability in a certain regime of operation. The underlying
mechanism is the slow adaptation of the processing rate to changes in the order flows or stock levels
in the market. Interestingly, the resulting oscillations in the inventories of the different products $i$ 
have a characteristic frequency, which can be much lower than the underlying fast variations in the consumption
rate. Depending on the network structure, 
these oscillations synchronize among different economic sectors and may explain business cycles
as a self-organized phenomenon with slow dynamics. These conclusions are not restricted to linear
supply chains, but can be generalized to many other supply networks, 
which are linearly unstable with respect to perturbations. In reality,
business cycles are, of course, less regular and of smaller amplitudes than in Figs.~\ref{FIG4} or \ref{Fig9}. However, 
the above model allows to reflect these aspects in a natural way
by inclusion of fluctuations, heterogeneity, additional
capacity constraints, and realistic network structures. Investigations with empirical data and for particular kinds of 
networks are on the way. It should also be noted that already deterministic
models of supply networks can show irregular, non-periodic behavior such as 
chaotic dynamics \cite{beer1,Chaos}, which calls for suitable control concepts \cite{control}.
\par 
Compared to traffic dynamics, economic and production systems have some 
interesting new features: Instead of a continuous space,
we have discrete production units, and the production speed as a function of the inventories is different from the empirical
velocity-density relation in traffic. Due to the minimum condition (\ref{mini}), production systems may
operate in different regimes, and small changes of parameters may have tremendous effects. 
For example, we may have a transition from small
oscillations of relatively high frequency to large oscillations of low frequency.  Apart from this, the management
strategies can vary to a large extent, and with this the control functions $W_b(\dots)$. With suitable strategies,
the oscillations can be mitigated or even suppressed \cite{Daganzo,Nagatani,Seba}. Moreover, production systems
are frequently supply networks with complex topologies rather
than linear supply chains, i.e. they have additional features compared
to (more or less) one-dimensional freeway traffic. They are more comparable to street networks of cities.
\par
Apart from some equations
which were not further applied in this study, most of the proposed model equations were 
conservation equations, equations given by the product flows, or relations derived 
with stochastic concepts used in queuing theory. They reflect the transport and interaction
of products, so that the physics of driven many-particle systems and of complex systems
can make some significant contributions to the new multidisciplinary field of  self-organization phenomena
in production and supply networks.
\par
Future work
will have to address questions such as the relevance of the network structure for the resulting dynamics,
possible control strategies, the role of the market and pricing mechanism, etc. This particularly concerns
the specification of the function $W_b$, and the equations suggested
in the paragraph on delivery networks and price dynamics. Here, we have already seen
that the supply network's topology and the level of heterogeneity in production systems
have a significant impact on the resulting dynamics: Parameter changes can
have tremendous effects. For example, a heterogeneity in the parameters characterizing  the
different production units can stabilize the production and the market
considerably, while in traffic flow, heterogeneity has normally a destabilization effect.
The stabilization of supply networks through heterogeneity probably could explain why the variations
in economic systems appear to be less dramatic than in our simulations, but additional inefficiencies
and capacity restrictions (corresponding to finite values of $S_b$, $I_b^{i,{\rm max}}$, and $O_b^{i,{\rm max}}$)
are probably another reason. This and the role of heterogeneity for the 
micro-macro link will be studied in more detail in the future. Should it turn out that the micro-macro link
requires a high degree of homogeneity in the parameters, it would be favourable 
to simulate economic dynamics based on a microscopic model of production and supply networks in the future.
The tendency for globalization certainly increases the degree of homogeneity, but it also tends to generate
larger oscillations in the inventories, i.e. more serious over- and underproduction. At least in some markets,
there are definite signs of a development in this direction. 
Forthcoming publications will, therefore, investigate alternative control strategies (including forecasts). 
First results on how to  decrease the instability of supply chains can be found in Refs.~\cite{Daganzo,Nagatani,Seba}.

\subsection{Advantages, Extensions, and Potential Applications}

As the variables in our dynamical model of supply and production networks are operational and
measurable, the model can be tested and calibrated with empirical data. Moreover, it
is flexible and easy to generalize. For this reason, it can be adapted to various applications. 
Our approach can be related to microscopic considerations such as
queueing theory or event-driven (Monte-Carlo) simulations of production processes, but, 
as it focusses on the average dynamics, it is numerically much 
more efficient and, therefore, suitable for on-line control.
Nevertheless, the formulas can be extended by noise terms to
reflect stochastic effects. Our system of coupled differential equations would
then become a coupled system of stochastic differential equations
(Langevin equations), where the noise amplitudes would be determined via
relationships from queueing theory.
\par
The dynamical theory of supply chains appears to be a promising field with many research opportunities.
It is not only useful for a deeper understanding of the origin and dynamics of business
cycles or for the optimization of production processes and supply
networks. It could also contribute to the further improvement of existing 
traffic control strategies in urban street networks or to the development of
more robust routing algorithms for internet traffic. Apart from this, the model can be viewed as a
dynamic multi-player game, where the individual control functions $W_b$ reflect the
strategies of the players $b$ in terms of quantities $N_i$, which represent the
information-feedback available to them. 
Generalizing this idea, the above model of supply networks may 
serve as a basis for particular kinds of neural networks. 
We are now setting up different projects in these directions, and cooperation is very welcome.

\subsection*{Acknowledgment} 

The author is grateful for inspiring discussions
with Arne Kesting, Megan Khoshyaran, Christian K\"uhnert, Stefan L\"ammer,
Tadeusz P\l{}atkowski, Dick Sanders, Thomas Seidel, 
Torsten Werner, and Ulrich Witt. He also likes to thank 
Tilo Grigat for preparing the schematic illustrations.

\end{document}